# Overview of Use Cases in Single Channel Full Duplex Techniques for Satellite Communication

*Victor Monzon Baeza[1], Steven Kisseleff [1], Jorge Luis González Rios[1], Juan Andrés Vasquez-Peralvo[1], Carlos Mosquera[2], Roberto López Valcarce[2], Tomás Ramírez Parracho[3], Pablo Losada Sanisidro[3], Juan Carlos Merlano Duncan[1], Symeon Chatzinotas[1]*

[1] Interdisciplinary Centre for Security, Reliability, and Trust (SnT), University of Luxembourg, 4365 Luxembourg, Luxembourg.
[2] atlanTTic Research Center, Universidade de Vigo, Galicia, Spain
[3] Gradiant, Vigo, Galicia, Spain
*email: victor.monzon@uni.lu

## Abstract

This paper provides an overview of the diverse range of applications and use cases for Single-Channel Full-Duplex (SCFD) techniques within the field of satellite communication. SCFD, allowing simultaneous transmission and reception on a single frequency channel, presents a transformative approach to enhancing satellite communication systems. We select eight potential use cases with the objective of highlighting the substantial potential of SCFD techniques in revolutionizing SatCom across a multitude of critical domains. In addition, preliminary results from the qualitative assessment are shown. This work is carried out within the European Space Agency (ESA) ongoing activity FDSAT: Single Channel Full Duplex Techniques for Satellite Communications.

## 1   Introduction

Over the past three decades, five successive generations of mobile communications have steadily improved connection speeds [1]. At the same time, a large percentage of our internet traffic is currently generated by mobile devices, which have become a necessity in the modern developed society. Despite this constant evolution in broadband speeds, each new cellular generation is deployed to a shrinking percentage of the coverage (mainly urban areas) to address the concentrated population, which constitutes a major percentage of the market [2]. With ever-increasing data volume demands accompanied by pressure to reduce the cost per bit delivered, new techniques to utilize the available radio frequency (RF) spectrum more efficiently are continually needed.

Satellite communication (SatCom) systems have traditionally been capable of providing the necessary coverage even to remote and under-served areas. Peak demands in the feeder and forward links can occur simultaneously and are difficult to satisfy. In this context, the full duplex operation (FD), consisting of simultaneous transmit and receive functionality applied to both links, seems especially promising.

Simultaneously transmitting and receiving information in a single frequency channel has already proven a viable approach in other wireless communication application domains [3-5] and has the potential to double total system throughput. Such systems typically employ a combination of both analogue and digital Self-Interference Cancellation (SIC) techniques, together with a careful antenna design and a suitable placement, to achieve the necessary isolation between transmit and receive signals, and such techniques have therefore advanced rapidly over the last few years [6-8]. An overview of channel models for SatCom can be found in [9]. However, those models have been considered the FD application, presenting a big challenge for SatCom.

We can consider two options for FD application, mainly we can refer to as In-Band Full-Duplex (IBFD) in general to simultaneously transmit and receive on the same channel, while Out-of-Band Full-Duplex (OBFD) counterparts' transmission and reception operate in nearby frequencies. IBFD can provide several benefits such as better spectral efficiency, reduced latency, or true channel reciprocity compared to OBFD. It has been traditionally avoided, except in very specific cases such as radar, RFID readers, or amplify and forward relays (repeaters), due to the high Self-Interference (SI) levels overwhelming the receiver. An appropriate SI cancellation (SIC) is instrumental for the operation of IBFD and, in general, for those cases with significant leakage from the transmit signal into the receive chain, which includes OBFD.

Single-channel full-duplex (SCFD) operation has consequently been supported within consumer (cable) networks since the release of DOCSIS 4.0 in 2017, and integrated wireless access and backhaul for 5G using full duplex is also now becoming a reality [10]. LTE products are already available using FD such as LTE User Equipment Relays, which enable a small cell LTE eNodeB to backhaul to an LTE macro base station using the same access frequencies for both the small cell and the macro network. That requires in most cases the use of one antenna, nowadays an optimized Direct Radiating Array (DRA) with beamforming capabilities [11] for transmission and another for reception, both working in a different frequency.

In most SatCom systems, FD operation is achieved using two separate frequency channels with spectral filtering to achieve the required isolation between signals in FDD. Although less widely utilized in SatCom, single-channel bi-directional communications are also sometimes realised by Time Division Duplexing (TDD), e.g., by the Iridium inter-satellite links (ISLs). Despite the recent advances, there remain significant barriers to applying full duplex operation within a SatCom system, chiefly due to the very high-power imbalance between the transmit and receive signals (a direct result of the very long transmission distances) and the large fractional bandwidths of SatCom frequency allocations [12-21]. In addition, we also highlight that the new flexible payloads, with onboard processing, allow improving regenerative functions, this will help in canceling interference to perform full-duplex operation [22].

In this context, this paper aims to identify and evaluate the possible use cases where it makes sense to use the SCFD techniques in satellite communications. The scenarios where these use cases can be applied will be proposed, as well as the proposal of a series of metrics and a preliminary assessment of said cases. This work is part of a European Space Agency project called FDSAT. Here we present preliminary results for the selection of potential use cases.

The rest of the paper is as follows: section 2 presents the use cases selected to use SCFD in SatCom. Section 3 performs a preliminary assessment of these use cases, followed by the conclusion in Section 4.

## 2. Potential Use cases (US) selected

A general SatCom system is shown in Fig. 1. Based on this scheme, various use cases can be found in the feeder (FL), user (UL), and control (CTRL) links as compiled in Table 1. These use cases are classified also for their application and frequency band targeted.

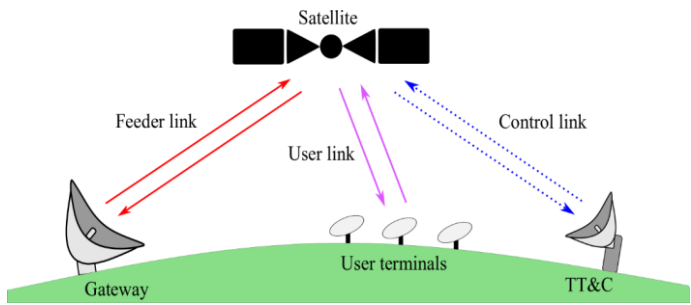

Fig. 1 Satellite communication system with feeder, user, and control links.

On the one hand, we consider the usage of the individual links. The FL is a link that concentrates a high amount of traffic, so it has high bandwidths that can be useful for broadband services. The current technologies are considering for FL to operate on Q/V and optical frequency bands. The UL use case involves 5G over satellite, hot spot scenarios, relaying, integrated access, and backhauling (IAB), satellite-aided M2M (Machine-to-Machine) and D2D (Device-to-Device) applications and can operate on Ku and Ka frequency bands. The CTRL use case is for the transmission of telemetry data to the satellite and can operate on C/Ku frequency bands.

On the other hand, we can consider the usage of combining different links to apply the FD operation. For example, the FU-UD use case involves broadcast, multicast, or unicast applications and can operate on Q/V, optical, Ku, and Ka frequency bands. The UU-FD use case is for backhauling scenarios.

Instead, the space-ground links, we can consider space-space links such as inter-satellite links (ISLs) and inter-orbital links (IOLs) as potential use cases for the FD operation. These links are used to exchange information among the satellites either of the same orbital plane (Single Orbit, SO) or among satellites belonging to different orbits in the recently envisaged NGSO satellite mega-constellations (multi-layer, ML).

Furthermore, space-air-ground integration has been suggested for the harmonization of future communications systems, such as 6G. The corresponding satellite-aerial links (SALs) connecting the satellites with the aerial vehicles, such as high-altitude platform systems (HAPSs) and unmanned aerial vehicles (UAVs) may benefit from a higher spectral efficiency enabled by FD, especially in the presence of many such links. In Fig. 2 we show the architecture for use cases space-space and space-air-ground links.

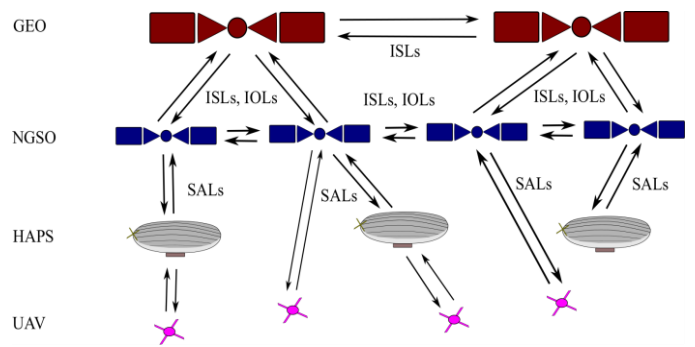

Fig. 2 Multi-layered space-air network with satellite-aerial links.

For these use cases, we highlight the advantages that make them potential for satellite communications in Table 2.

## 3 Use cases assessment

The presented UCs are qualitatively and quantitatively analysed from the perspective of potential benefits and relevance for future high-throughput Satcom systems.

Table 1 Identified use cases and their applications.

| ID | Use case | Application | Frequency band |
|---|---|---|---|
| UL | User link | 5G over satellite, hot spot scenarios, relaying, integrated access, and backhauling (IAB), satellite-aided M2M and D2D | Ku, Ka |
| FL | Feeder link | Broadband | Q/V, optical |
| CTRL | Control links | Transmission of telemetry data to the satellite | C/Ku |
| FU-UD | Feeder uplink and user downlink | Broadcast/multicast/unicast. | Q/V, optical, Ku, Ka |
| UU-FD | User uplink and feeder downlink | Backhauling | Q/V, optical, Ku, Ka |
| ISL-SO | Inter-Satellite Link (ISL) for a single orbital plane | Relaying, data offloading, exchange of parameters, e.g., telemetry, machine learning model, etc. | L/S/Ka/W, optical |
| ISL-ML | Inter-satellite and inter-orbital links in multi-layer satellite constellations | Relaying, data offloading, edge computing, exchange of parameters, e.g., telemetry, machine learning model, etc. | L/S/Ka/W, optical |
| SATL | Satellite-aerial-terrestrial links | Satellite-aerial terrestrial integration, data offloading, edge computing | K |

Table 2 Advantages for the identified use cases.

| ID | Advantages |
|---|---|
| UL | Doubling the spectrum (increasing data rate or number of served users), reducing latency by up to 50%, reducing power imbalance between terrestrial and satellite networks, and reducing interference between the bands. |
| FL | Doubling the spectrum (especially useful for feeder links. This can help to reduce the required number of feeder links to satisfy the demand), reducing interference between adjacent bands, and reducing latency by up to 50%. |
| CTRL | Reduction of the orbital arc between adjacent satellites, and reduction of latency by up to 50%. |
| FU-UD | Reduction of the interference between very close bands, overhead reduction, doubling of spectrum. |
| UU-FD | Reduction of the interference between very close bands, overhead reduction, doubling of spectrum. |
| ISL-SO | Doubling the spectrum, reducing the power imbalance between terrestrial and satellite networks, reduction of latency by up to 50%. |
| ISL-ML | Doubling the spectrum (e.g., to increase the number of simultaneously inter-connected satellites), reducing the orbital arc between the satellites, reduction of power imbalance between terrestrial and satellite networks, reduction of latency by up to 50%. |
| SATL | Doubling the spectrum (e.g. increasing the number of supported airplanes and HAPS), reducing of power imbalance between terrestrial and satellite networks |

3.1 Qualitative assessment

- **ISL-SO/ISL-ML**: are the most promising use cases in terms of compliance with the current regulations. Moreover, the information transmission over ISLs can substantially reduce the load and relax the requirements for terrestrial systems. Given the variety of possible applications, such as data offloading and exchange of parameters for joint optimization, the extensive use of ISLs is envisioned in future Satcom systems, which makes these use cases especially relevant. Also, regarding the latency of communication, ISLs have been demonstrated to outperform fiber-optical connections over large transmission distances. The benefit of full duplex operation in this context will thus manifest in shifting the balance between terrestrial and satellite communication by improving the connectivity among the satellites.
- **FL/UL** are very promising in terms of advantages provided by the full duplex operation, which can especially help to reduce the load on the respective links. In fact, both feeder and user links are expected to be heavily loaded in the future Satcom systems due to the ever-growing demand for Satcom services. Accordingly, the benefits of full-duplex technique in terms of spectrum doubling would substantially increase the Satcom network throughput. In addition, M2M and D2D communication via satellite relay can be enhanced using full duplex, if the ground users for the downlink and uplink are different. Similarly, the satellite can be connected to different gateways in the feeder uplink and downlink. This would improve the balancing of the load. Furthermore, IAB can be implemented to provide simultaneous access and backhauling to the ground users.
- **SATL:** While the satellite-aerial-terrestrial integration is envisioned in principle, the amount of signalling that comes along with this integration might be substantial and can drastically impact the overall performance. Accordingly, the benefits of full duplex for the whole system with this use case are the result of a trade-off between coordination complexity, signal quality, and available spectrum.
- **CTRL:** The full-duplex operation over the control link is very advantageous as it allows to improve the control over the spacecraft by reducing the latency. However, the low latency of control information is especially beneficial in the case of maneuvering the satellite on NGSO orbits. Since permanent maneuvering does not seem to be feasible (unless in scenarios with satellite swarms), this use case is not prioritized.
- **FU-UD/UU-FD:** These use cases are interesting from the perspective of their application, I.e., backhauling, broadcasting, etc. However, since feeder links and user links typically operate in different frequency bands, it may not be feasible to employ full duplex. In fact, a change of regulations may be required in this case, which can take a long time to enforce.

3.2 Quantitative assessment

For the quantitative analysis, we have selected the mixed case of feeder link and user link, as well as the case of SATL, to obtain preliminary results of the analysis. We consider a LEO satellite constellation based on Iridium as an example whose parameters are:

- An altitude of 780 km,
- 11 satellites per plane, and
- 6 orbits.

The rest of the parameters for the preliminary study are:

- Minimum elevation angle for ground stations is 10 degrees.
- TT&C and Gateway (GWs) stations are assumed to be deployed at the University of Luxembourg (latitude 49.6266N, longitude 6.15898E).
- User terminals are at the University of Vigo (latitude 42.16951N, longitude 8.68318W).
- Equivalent Isotropic Radiated Power (EIRP) of 43 dBW for user terminals, 65 dBW for satellite and 43 dBW for ground stations as Gateway.
- G/T value of 31.5 dB/K The frequency selected for evaluating the FD operation in IBFD per each use case is the central frequency in the range defined by the band.

Regarding carrier frequencies, we select the following depending on the link considered in each use case:

- For user links: 29.3 GHz (Ka Band).
- For feeder links: 37.5 GHz (Q/V Band).

The isolation level values considered in circulators are obtained from [23] according to the carrier frequency. These values correspond to 25 dB in user equipment or satellite/HAPS and 40 dB in ground stations.

In addition, for the signal bandwidth, we assume 50 MHz. The signal amplification at the satellite is set to 60 dB. Also, we assume a temperature of 290K in all receivers. We propose to assume 70 dB of SIC for all use cases as an example. Then, it is possible to evaluate the performance gain compared to the benchmark, which is half-duplex (FDD) with the same total bandwidth.

To show the preliminary results we have selected the case that considers both FL and UL. In the FU-UD case, the FD application is considered only in the feeder, while for UU-DD, the full-duplex is considered for the user uplink.

3.2.1. FU-UD use case

For the SIC objective, we obtain the following spectral efficiency (SE) of 6.43 bps/Hz with half-duplex (FDD) and 10.74 bps/Hz in FD. Here, we obtain a gain of approx. 67%. By enhancing the SIC level to 80 dB, the gain can be further increased by 15%.

3.2.2. UU-FD use case

The SE obtained is 7.21 bps/Hz with half-duplex (FDD) and 11.26 bps/Hz with FD. Here, we obtain a gain of approx. 56%. By enhancing the SIC level to 80 dB, the gain can be further increased by 24%.

3.2.3. SATL use case

We obtain the following a SE of 3.26 bps/Hz with half-duplex (FDD) and 5.52 bps/Hz with full duplex. Here, we obtain a gain of approx. 70%. By enhancing the SIC level to 80 dB, no significant gain is observed.

# 4 Conclusion

This contribution focuses on the identification of promising use cases in the SatCom domain. The most promising use cases in this analysis are ISL-SO/ISL-ML and FL/UL, while less promising SATL, CTRL, FU-UD, and UU-FD. ISL-SO and ISL-ML are the most promising use cases in terms of compliance with the current regulations. FL/UL are very promising in terms of advantages provided by the full duplex operation, which can especially help to reduce the load on the respective links. We assess the advantages and remark for different cases. In addition, a preliminary evaluation for each use case is shown for a target SIC value. AS future work, we extend the analysis for GEO and several LEO constellations.

## 5 Acknowledgements

This work has been supported by European Space Agency (ESA) funded ongoing activity FDSAT: Single Channel Full Duplex Techniques for Satellite Communications, contract 4000140117/22/NL/CLP). The views of the authors of this paper do not necessarily reflect the views of ESA.